     \documentclass{article} \catcode`\@=11
     \usepackage{rotate,epsf}
    \def\section{\@startsection{section}{1}{\z@}%
    {-3.5ex plus -1ex minus -.5ex}{1.5ex plus.3ex}{\bf }}
    \def\subsection{\@startsection{subsection}{1}{\z@}%
    {-3.5ex plus-1ex minus-.5ex}{1.5ex plus.3ex}{\bf }} 
    
\begin{document}
    \hfill\parbox{4.77cm}{\Large\centering Annalen\\der
    Physik\\[-.2\baselineskip] {\small \underline{\copyright\ Johann
    Ambrosius Barth 1998}}} \vspace{.75cm}\newline{\Large\bf
Critical level statistics at the Anderson transition in 
four-dimensional disordered systems
    }\vspace{.4cm}\newline{\bf   
I.Kh.\ Zharekeshev and B.\ Kramer
    }\vspace{.4cm}\newline\small
I. Institut f\"ur Theoretische Physik, 
Univesit\"at Hamburg,
Jungiusstra{\ss}e 9, 
D-20355 Hamburg, Germany
    \vspace{.2cm}\newline 
Received 6 October 1998
    \vspace{.4cm}\newline\begin{minipage}[h]{\textwidth}\baselineskip=10pt
    {\bf  Abstract.}
The level spacing distribution 
is numerically calculated 
at the disorder-induced metal--insulator transition
for dimensionality $d=4$ by applying the Lanczos diagonalisation. 
The critical level statistics are 
shown to deviate stronger from the result
of the random matrix theory compared to those of $d=3$
and to become closer to the Poisson limit of uncorrelated spectra.  
Using the finite size scaling analysis for the probability distribution
$Q_n(E)$ of having $n$ levels in a given energy interval $E$
we find 
the critical disorder $W_{c}=34.5\pm 0.5$,
the correlation length exponent $\nu=1.1 \pm 0.2$ and
the critical spectral compressibility $\kappa_c\approx0.5$.
    \end{minipage}\vspace{.4cm} \newline {\bf  Keywords:}
Localisation; Metal-insulator transition; 
Level statistics; Critical phenomena
    \newline\vspace{.2cm} \normalsize

\section{Introduction}
The statistical description of electronic spectra 
at the Anderson transition 
is one of the subjects of central interest. 
This quantum phase transition 
implies that
there exists a non-vanishing value of the disorder of the random potential, 
at which the system undergoes the crossover between a conducting 
and an insulating phases
with delocalised and localised electron states, respectively.
The spatial dimensionality $d$ of the system 
plays an important role in determining the critical properties 
of both the conductivity and the level statistics.
According to the one-parameter scaling theory of 
localisation~\cite{AALR79}  
all single-electron states in the one and two dimensions 
are localised even for arbitrarily weak randomness, 
provided that the time-reversal and the spin-rotational
symmetries are preserved. Therefore,
the lowest integer dimension for which the disorder-induced metal-insulator 
transition (MIT) occurs for non-interacting particles is $d=3$.
Numerous computer simulations performed on transport electron properties
have confirmed the scaling hypothesis
(see, for example, \cite{MackinnonK81,KramerM93} 
and references therein). 
Furthermore, 
the MIT 
has been also found at the next higher integer
dimensionality $d=4$,
by applying the transfer-matrix method (TMM)~\cite{MarkosH94,SchreiberG96}.

The energy level statistics in two dimensions (2D)
do not exhibit critical 
behaviour~\cite{Shklovskii93,ZharekeshevBK96},
while in 3D they do 
\cite{Shklovskii93,HofstetterS94,Evangelou94,ZharekeshevK94}
in agreement with the earlier results obtained 
by the TMM~\cite{KramerM93}.
A new universal scale-independent
level spacing distribution $P(s)$ has been predicted 
exactly at the MIT
in 3D~\cite{Shklovskii93}. 
This distribution 
differs characteristically
from the results of the `classical' random matrix theory (RMT)
by Wigner and Dyson~\cite{Wigner55,RMT}, which is valid for weakly
disordered conductors~\cite{Efetov83,AltshulerS86}. 
An analytical approach has been 
developed recently~\cite{AronovKL94,Kravtsov94}, 
which supposes that the shape of the 
$P(s)$ at the mobility edge is given by a combination of the 
dimensionality $d$ and 
the critical exponent $\nu$ of the correlation length. 
At present this relation is intensively investigated for models
of different basic symmetry, namely for
the orthogonal~\cite{ZharekeshevKJ95,HofstetterV95,ZharekeshevK97}, 
the unitary~\cite{HofstetterS95,BatschLZ96}
and the symplectic~\cite{SchweitzerZ95,Evangelou95,Kawa96} critical 
ensembles.
Here the critical disorders and the exponents are known, and the
dependence of $P(s)$ on $\nu$ and $d$ could be determined quantitatively.
In order to obtain further insight,
it is of great interest to study also critical parameters and
the level statistics at higher dimensions.

In this paper we report results of detailed numerical calculations 
on the level statistics using 
the Anderson model for the 4D hypercubic lattice. 
We show that the nearest-neighbour level spacing distribution $P(s)$
exhibits critical behaviour. Its shape is {\em size-invariant} 
at the transition, and its form
considerably differs from that in 3D.
Using the level statistics method in terms of the complete distribution
of having $n$ levels in a given energy interval 
we detect the metal-insulator transition,
which corresponds to the disorder $W_{c}=34.5\pm~0.5$.
We perform the finite size scaling analysis in order to find the
one-parameter scaling function and to determine 
the correlation length exponent $\nu$, which is smaller than that
in 3D.
Combining the results obtained for the statistic $J_0$ (defined below)
and the variance of the level number
we argue that 
the {\it {critical}} spectral fluctuations are
stronger in 4D than in 3D, being closer to the Poissonian statistics.

\section{Model and computational procedure}
The model for the 4D disordered system
is defined by the Anderson Hamiltonian
\begin{equation}
H=\sum_{n}\epsilon_{n}^{} c_{n}^{\dag} c_{n}^{} +
       V \sum_{n \neq m} (c_{n}^{\dag} c_{m}^{} + c_{n}^{} c_{m}^{\dag}),
\label{Hamil}
\end{equation}
where $c_{n}^{\dag}$ ($c_{n}^{}$) is the creation 
(annihilation) operator of
an electron at a lattice site $n$,
and $m$ denotes the sites adjacent to the site $n$
(their number equals 8 for the simple hypercubic lattice).
The site energies $\epsilon_{n}$ 
are randomly distributed according to a box distribution with 
a width $W$, which plays the role of the disorder parameter.
The second term
describes the hopping between the nearest-neighbour sites in the lattice.
Our considerations are restricted to the particles without spin and
with no magnetic field.  
It was earlier found by the TMM~\cite{MarkosH94} that
the MIT 
at the band centre $E$=0
is close to the disorder $W=33.2$. 
For 3D the critical disorder is smaller, 
$W \approx 16.5$~\cite{MackinnonK81}.

After numerical diagonalisation of the Hamiltonian~(\ref{Hamil})
with periodic boundary conditions
using the Lanczos algorithm we obtained the exact discrete spectrum 
of the electrons for simple 4D hypercubic lattices of various sizes
ranging from $L^4=4^4$ to $10^4$ and for different disorders~$W$.
Linear and energy scales are measured in units of 
the lattice constant and the overlap integral between adjacent sites
($V=1$), respectively. 
The levels were taken from energy intervals centred at $E=0$
so that they belong to the critical energy region, 
defined by the condition $L<\xi = (|E-E_{c}|/E_{c})^{-\nu}$, 
where $\xi$ is the correlation length and $E_c$ is the mobility edge.
The number of realizations for a given size $L$ 
was such that the total number of eigenvalues amounted as much as $10^5$.
We have checked that the density of states $\rho = (\Delta L^{4})^{-1}$
around the band centre slightly varies with the energy
 ($\Delta$ is the mean level spacing). Therefore, the careful
unfolding procedure has been applied for the 
spectra of all pairs of \{$W,L$\}.

\section{The critical level spacing distribution $P_c(s)$}
A traditional way to study the statistical properties of disordered spectra 
is to consider the level spacing distribution $P(s)$ which is defined as 
the probability density of nearest neighbouring levels.
It is known that in the metallic region $P(s)$ is very close
to the Wigner surmise~\cite{Efetov83}
for the Gaussian orthogonal ensemble (GOE)
of random matrices~\cite{Wigner55}, namely 
\begin{equation}
P_{\rm GOE}(s)=\frac{\pi}{2}\,s\,\exp\left( -\frac{\pi}{4} s^2 \right ),
\label{GOE}
\end{equation}
where $s$ is measured in units of the mean level spacing~$\Delta$. 
In the localised region the energy levels are 
completely uncorrelated, and hence 
the spacings are distributed according to the
Poisson law 
\begin{equation}
P_{\rm P}(s) = \exp(-s).
\label{Poisson}
\end{equation}
In a similar way as for the 3D Anderson 
model~\cite{Shklovskii93,HofstetterS94,ZharekeshevK94}, 
the third, universal distribution $P_{c}(s)$
is supposed to be revealed exactly at the critical 
point, 
which is different from both of the above  
laws.
For systems of finite size the level statistics is expected 
to change continuously
from $P_{\rm GOE}(s)$ through $P_{c}(s)$ to $P_{\rm P}(s)$, 
when increasing the strength of fluctuations of the
 random impurity potential.
When increasing the size of the system the distribution 
tends towards
either $P_{\rm GOE}(s)$~(\ref{GOE}) or $P_{\rm P}(s)$~(\ref{Poisson}), 
depending on whether the
disorder is below or above its critical value, respectively.

\begin{figure}[t]
\hspace*{10mm}
\hbox{\rotate[r]{
{\epsfysize=95mm\epsfbox{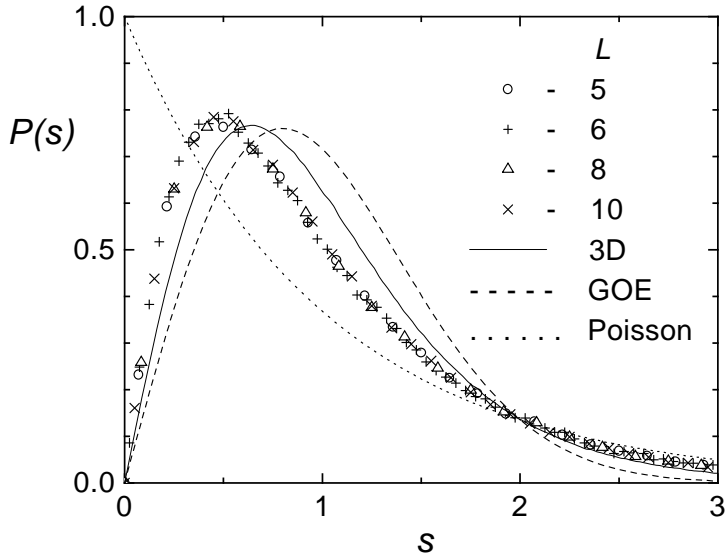}}
}}
\caption[]{\protect\small The nearest neighbour 
level spacing distribution $P_c(s)$ for $d=4$ 
at the disorder $W=34.5$ 
for different lattice sizes~$L^4$.  
Full line is $P_{c}(s)$ for $d=3$ taken from~\cite{ZharekeshevK97}.
Dashed and dotted curves are $P_{\rm GOE}(s)$~(\ref{GOE}) 
and $P_{\rm P}(s)$ ~(\ref{Poisson}),
the distributions for the metallic and insulating limits, 
respectively.}
\label{fig1}
\end{figure}

Fig.~\ref{fig1}\, shows the function $P(s)$ calculated 
at the  disorder $W$ very close to  the critical point.
The fact that 
the data within numerical errorbars 
lie on a common curve 
independent on $L$,  defines the critical 
level spacing distribution 
$P_c(s)$ 
(see also section~5).
In comparison with $d=3$ the critical $P_{c}(s)$ for $d=4$ is closer 
to $P_{\rm P}(s)$.
It is worth noticing that our numerical data deviate considerably from
the interpolation formula $P_c(s)=B\,s \exp(-A\,s^{1+1/d\nu})$ derived
analytically in~\cite{Kravtsov94}, where the coefficients $A$ and $B$
are defined by the normalisation conditions.

\begin{figure}[t]
\hspace*{17mm}
\hbox{\rotate[r]{
  {\epsfysize=80mm\epsfbox{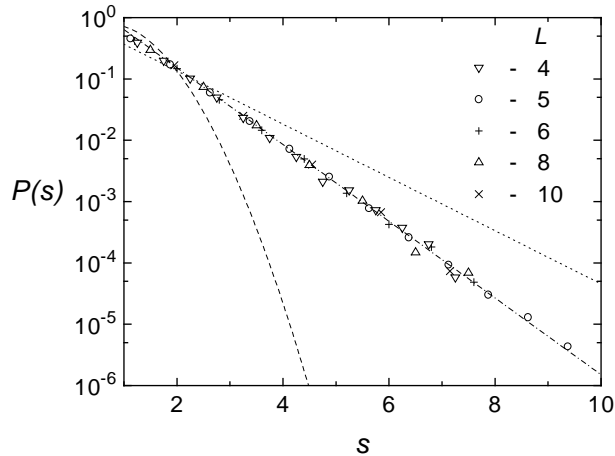}}
  }}
\caption[]{\protect\small 
Large-$s$ part of the critical level spacing distribution 
$P_c(s)$ of the Fig. 1 for various system sizes~$L$.
Dashed-dotted straight line, $\ln P_c(s)=-1.4 s$, 
is the best fit to the data.
Dashed and dotted curves are $P_{\rm GOE}(s)$ 
and $P_{\rm P}(s)$, respectively.}
\label{fig2}
\end{figure}

For small spacings we found the linear increase  $P_{c}(s)=B s$, 
as consistent with the orthogonal symmetry.
However, the linear slope $B\approx 2.1 B_{\rm GOE}$ is larger than in 
$d=3$~\cite{Shklovskii93} 
and, consequently, in the RMT. This indicates 
that the level repulsion becomes weaker with increasing~$d$.
 The behaviour of $P_{c}(s)$ at larger
spacings 
is well described by the sub-Poissonian form 
$P_c(s) \propto \exp(- A s)$ with $A = 1.4 \pm~0.1$.
One can see in 
Fig.~\ref{fig2} that $P_c(s)$ changes by several orders of magnitude
in the interval $2<s<10$.
As expected, this asymptotic decay is slower compared to $d=3$, 
where the exponential rate is $A \approx 1.9$~\cite{ZharekeshevK97},
but faster than the Poissonian decay.

\section{The probability distribution  $Q_{n}(s)$}
In order to perform the finite size scaling analysis and to study
how the level statistics behave around the critical point, we have
calculated the dependence of the statistics of neighbouring spacing on
the disorder $W$ for different~$L$.
To include the entire range of spacings, 
we use here 
the probability distribution of having $n$ eigenvalues in a given energy
interval of the width~$s$: 
\begin{equation} 
Q_{n}(s) \equiv \int_{s}^\infty I_{n}(s^\prime)ds^\prime 
=\int_s^\infty ds^\prime \int_{s^\prime}^\infty p_{n}(s^{\prime\prime}),
ds^{\prime\prime},
\end{equation}
where $I_n(s)$ being the cumulative distribution of $n$ successive levels.
The function $Q_n(s)$ is known for the metallic limit
from the RMT~\cite{RMT}. In the insulating limit the Poisson process 
$Q_{n}(s)=s^n \exp (-s)/ n!$ governs the completely uncorrelated spectrum. 

\begin{figure}[t]
\begin{picture}(120,192)
\put(-40,-145)
{\epsfysize=192mm\epsfbox{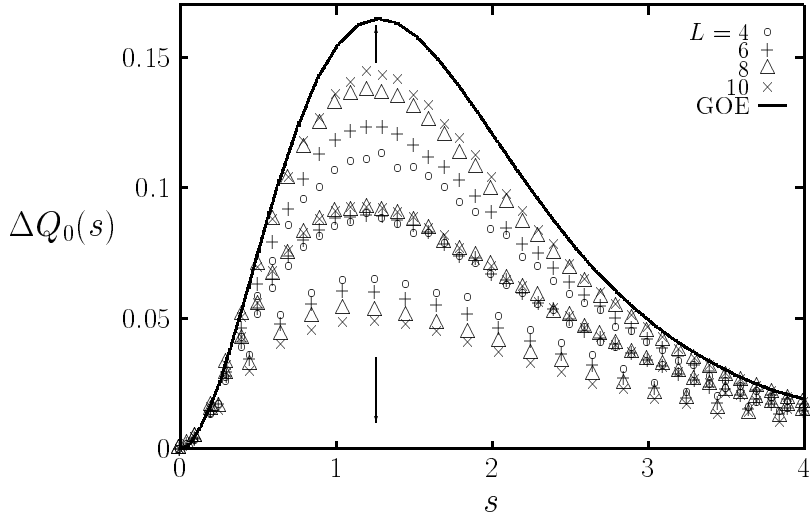}}
\end{picture}
\caption[]{\protect\small The probability distribution $\Delta Q_{0}(s)$ 
at different disorders $W=20$ (upper set), 33.2 (middle set) and 40 
(lower set) for various sizes $L$ of 4D hypercubic lattice.
Solid line is the RMT result taken from the Table A-19 of Ref.~\cite{RMT}.
The arrows indicate the direction of the size effect on either side
of the transition.}
\label{fig3}
\end{figure}

In what follows we investigate the probability to have no level, $n=0$,
within the bin $s$, which implies the level statistics for the case 
of nearest neighbour spacings.
In order to show ($L,W$)-dependence 
we plot in Fig.~\ref{fig3} the absolute deviation from the Poisson process
$\Delta Q_{0}(s) = -[Q_{0}(s) - \exp(-s)]$.
The data for the metallic phase, i.e for $W<W_{c}$,
turn out to be closer to the GOE. The larger the system size, the 
closer the level statistics to this limit. 
When $W>W_{c}$ the data approach zero, 
however with the opposite size effect. 
Closer to the critical disorder all of the data within statistical 
uncertainties start to fall onto
a common intermediate curve independent of~$L$. 
This is the manifestation of 
critical behaviour. 
Similar scaling properties are also observed for $Q_{n > 0}(s)$.

\subsection{The spectral statistic $J_n$}
To parameterise the distribution $Q_{n}(s)$ 
we deal with a global statistical quantity defined 
as follows 
\begin{equation}
J_{n} \equiv  \int_{0}^\infty Q_{n}(s) ds,
\label{Iofnot}
\end{equation}
which takes the whole range of the spacings 
into account.
For $n=0$ one can easily show that it is related to the spacing
variance $\langle \delta^2 s \rangle$ as
\begin{equation}
J_0= \frac{1}{2}\langle s^2 \rangle  = 
 \frac{1}{2} (\langle \delta^2 s \rangle + 1).
\label{varspa}
\end{equation}
For investigating the critical properties of the spectra at the
Anderson transition the statistic $J_{0}$  
was for the first time introduced in~\cite{ZharekeshevKJ95}. 
It proved to be more efficient than those which have previously 
been used in studying the 2D and 
3D cases~\cite{Shklovskii93,HofstetterS94,ZharekeshevK94}.
This is due to that since the probability density $P(s)=p_0(s)$ 
and the cumulative distribution $I_{0}(s)$
are normalised to unity, 
one had to
choose  some spacing $s^{\star}$ 
in order to weigh the functions 
after or before this point
~\cite{Shklovskii93,HofstetterS94,ZharekeshevK94}
for a given pair of the parameters \{$W,L$\}.
This leads  however to loosing part of the information
obtained from the  diagonalisation and, as a result, 
decreases the accuracy.
In contrast, the set of parameters $J_n$ 
does not require~$s^\star$. 
Therefore,
$J_n$ is  more successful,
in our opinion, 
for  demonstrating the scaling properties of the level statistics
with less number of realizations 
and, as consequence, to locate the MIT more 
precisely.
For the GOE 
$J_{0}=0.643$, $J_{1}=0.922$,~...,~$J_{n=\infty}=1$~\cite{RMT}.
In the localised regime, 
one has  simply $J_{n}=1$ for any~$n$. 
For $d=3$ the set of the critical numbers
$J_{n}^{c}$ has been obtained in~\cite{ZharekeshevKJ95}, for example,
$J_{0}^{c}\approx 0.714$ and $\langle s^2 \rangle \approx 1.42$.
In the presence of strong spin-orbit interactions, where the MIT
occurs even for $d=2$, another set $J_{n}^{c}$ describes 
the critical symplectic ensemble~\cite{SchweitzerZ95}. 

\begin{figure}[t]
\hspace*{12mm}
\hbox{\rotate[r]{
  {\epsfysize=95mm\epsfbox{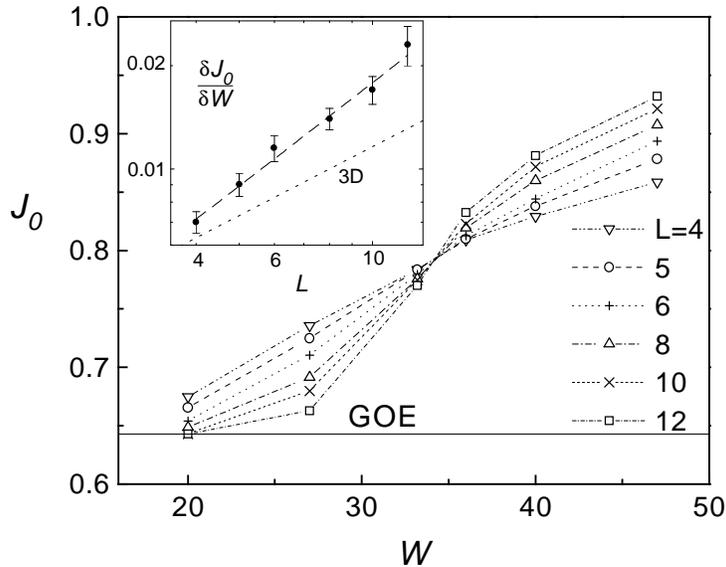}}
  }}
\caption[]{\protect\small 
Disorder dependence of $J_{0}$ for various sizes $L$.
Full line is the RMT result $J_0^{\rm GOE}=0.643$~\cite{RMT}. 
Inset: the derivative $dJ_{0}/dW$ near $W_{c}$ vs. size $L$.
Fit of numerical data shown by dashed line yields $\nu=1.1$.
For comparison, the slope of dotted line corresponds schematically 
to $\nu=1.45$ for $d=3$~\cite{ZharekeshevK94}.}
\label{fig4}
\end{figure}

\subsection{The finite size scaling and the critical exponent}
We have calculated the dependence of $J_{0}$ on $W$ near the transition
for various system sizes. 
All computed data lie within the interval between 
$J_{0}^{\rm GOE}$ and $J_{0}^{\rm P}$, gradually growing from 
the former to the
latter limit, when increasing the disorder~$W$.   
The increase of $J_{0}$ develops faster with $L$.
For an infinite system this change would transform to  a discontinuous
crossover between these two limits exactly at the transition $W_c$.
 One observes from Fig.~\ref{fig4}, that  $J_{0}(W)$ 
exhibits critical behaviour.
The common crossing point $J_{0}^c \approx 0.79$ 
($\langle s^2 \rangle \approx 1.57$), 
where the size effect 
on the statistics changes sign, corresponds to the transition. 
Intersection of the curves of different $L$
enables us to determine the fixed point more precisely, 
$W=W_c=34.5 \pm 0.5$. The obtained value is
 in reasonable agreement with that computed previously by the 
TMM~\cite{MarkosH94,SchreiberG96}. 
However it markedly deviates from
the linear relation $W_c(d)= (d-2) W_c(d=3)$~\cite{MarkosH94}. 

At the fixed point the 
correlation length diverges with the exponent $\nu$:
\begin{equation} 
\xi(W) \propto |W - W_c|^{-\nu}.
\label{local}
\end{equation} 
By using the lowest terms of the expansion
for the function $J_{0}(W,L)$ near $W_{c}$
\begin{equation} 
J_{0}(W,L) \approx J_{0}^c + C (W-W_{c}) L^{1/\nu} = 
J_{0}^c + C \left(\frac{L}{\xi} \right )^{1/\nu},
\label{expan}
\end{equation} 
one extracts the critical exponent of the correlation length. 
Inset of Fig.~\ref{fig3} shows that the data within the 
numerical errors are 
well described by the linear approximation
\begin{equation}
\ln \frac{dJ_{0}(W,L)}{dW} \propto \nu^{-1} \ln L.
\label{deriv}
\end{equation} 
The estimated value $\nu=1.1 \pm 0.2$ is consistent with previous
findings~\cite{SchreiberG96}. It appeared to be smaller
 than that for $d=3$ 
($\nu \approx 1.45$~\cite{ZharekeshevK94}). On the other hand, it is
still larger than the standard mean-field result
$\nu_{\rm MF}=1/2$,
valid for the upper bound $d_u$ 
of the Anderson transition~\cite{Efetov90}. 
Based on our results one can argue that $d=4$ is definitely lower
than $d_u$, which is believed to be equal infinity~\cite{Mirlin94}. 

\begin{figure}[t]
\begin{picture}(120,175)
\put(-26,-165)
  {\epsfysize=230mm\epsfbox{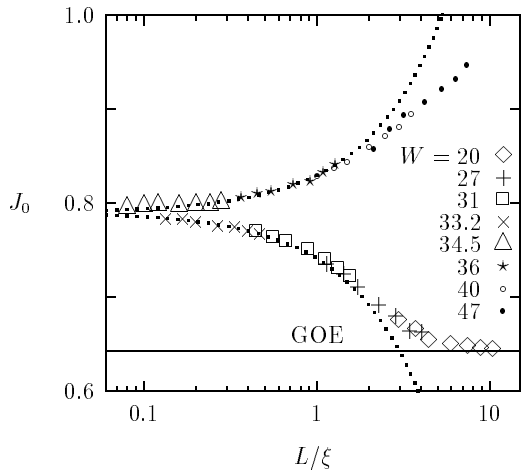}}
\end{picture}
\caption[]{\protect\small 
One-parameter scaling dependence of the statistic $J_{0}$
on $L/\xi$ for different system sizes $L$ and disorders~$W$. 
Full straight line is the RMT result $J_0=0.643$~\cite{RMT}.
Dotted line is~(\ref{expan}) with $J_0^c=0.79$ 
and the coefficient $C=0.043$.}
\label{fig5}
\end{figure}

\begin{figure}[t]
\begin{picture}(120,150)
\put(0,-150)
  {\epsfysize=200mm\epsfbox{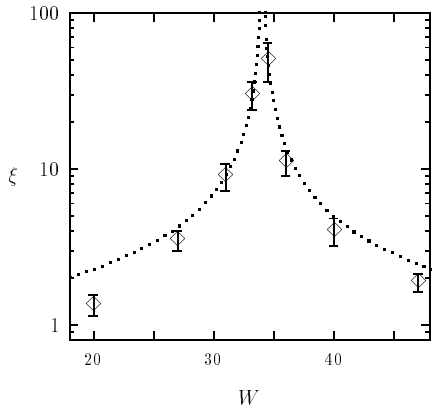}}
\end{picture}
\caption[]{\protect\small 
Disorder dependence of the correlation length $\xi(W)$
for 4D disordered system.
Dotted line is~(\ref{local}).}
\label{fig6}
\end{figure}

By introducing the scaling variable $\xi$, which is identified as
the correlation length, it is possible to replot all of the data 
in Fig.~\ref{fig4} 
into a single-parameter function, as shown in
Fig.~\ref{fig5}. The resulting curve $J_0(L/\xi)$ 
consists of two branches characteristic
of different regimes of the disorder-induced MIT.
 The decaying branch corresponds to the metallic phase ($W<W_c$) 
and the  growing one belongs to the insulating phase ($W>W_c$).
Outside of the critical region, i.e. $L/\xi > 1$, the numerical 
data deviate from the linear approximation~(\ref{expan}).
The one-parameter finite size scaling procedure allows one to find
the dependence of the correlation length $\xi$ on the disorder.
Close to $W_c$ the numerical results of $\xi(W)$ shown in Fig.~\ref{fig6}
give a satisfactory agreement with~(\ref{local}), 
while far apart from the fixed point one observes a considerable
discrepancy.
Thus, the critical behaviour of the spectral statistics in 4D 
is typical for the Anderson transition and analogous to that
for the 3D case~\cite{Shklovskii93,HofstetterS94,ZharekeshevK94}.

\section{Variance of the energy level number}
While the spacing distribution $P(s)$ and, consequently, $J_{0}$ probe 
the short-range correlations in the electron spectra, 
the variance of the number of levels 
$\langle \delta^2 N(E) \rangle$ in a given energy interval $E$,
which describes the global spectral rigidity,
can provide the information about fluctuations on scales
much larger than~$\Delta$. It is defined as a width of the distribution
of $N$ levels in the interval $E$: 
$\langle \delta^2 N(E) \rangle 
=\sum_{N=0}^{\infty}(N-\langle N \rangle) Q_N(E)$. 

In the extreme insulating limit $W \gg W_c$
the number variance obeys the ordinary Poisson law 
$\langle \delta^2 N(E)  \rangle=\langle N(E)\rangle$. For any finite
$W$ the eigenfunctions can spatially overlap, so that the fluctuations
$\langle \delta^2 N \rangle$ are reduced below the Poisson limit due to
level repulsion. 
When the system is a good conductor, i.e $W\ll W_c$, the electron states
are spreaded over the entire volume. Therefore the number variance 
for $\langle N \rangle \gg 1$ can be approximated 
by the Dyson formula
\begin{equation}
\langle \delta^2 N \rangle = \frac{2}{\pi^2} 
\ln \langle N \rangle + \gamma, \qquad
\gamma \approx 0.44,
\label{Dyson}
\end{equation}
valid for the Gaussian orthogonal ensemble of random 
matrices~\cite{RMT,Efetov83}.  
Thus, the relative variance 
$\langle \delta^2 N \rangle /\langle N \rangle$
for large $\langle N \rangle$, which is also known as a spectral
compressibility $\kappa$,
changes from zero to unity as $W$ increases. Of particular interest is
the question how does the relative variance behave at $W=W_c$. 
For instance, in 3D systems it was numerically shown that 
its critical value
equals a constant $\kappa_c \approx 0.27 $~\cite{ZharekeshevKJ95}.
It may also be worth considering
 in more detail 
the number variance for weakly disordered 4D systems in order to compare
with the results of the diffusive theory~\cite{AltshulerS86}.

\begin{figure}[t]
\hspace*{24mm}
{\epsfysize=6cm\epsfbox{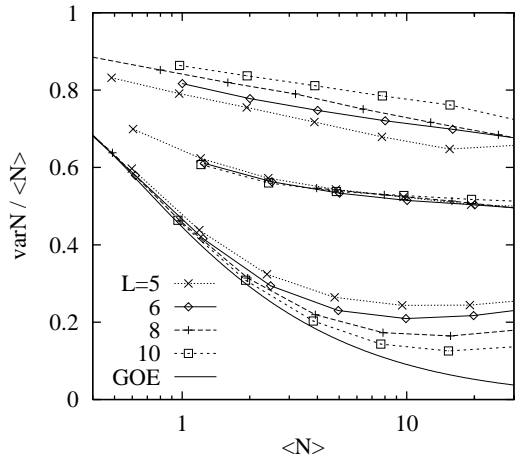}}
\caption[]{\protect\small 
The relative number variance 
$\langle \delta^2 N \rangle /\langle N \rangle$ of the spectrum 
as a function of the
mean number of levels in a given energy interval $\langle N(E) \rangle$
for various sizes $L$ at three disorder values: $W=20$ (lower set),
34.5 (central set) and 47 (upper set). 
Full line corresponds to~(\ref{Dyson}).}
\label{fig7}
\end{figure}

Fig.~\ref{fig7} shows the numerical results 
of the relative number variance around the transition.
The size dependence below and above the critical point is very similar
to that of $\Delta Q_0(s)$ in Fig.~\ref{fig3}. With increasing $L$ 
the data approach the Dyson result~(\ref{Dyson}) if $W<W_c$, while 
$\kappa \to 1$ if $W>W_c$, respectively. 
At $W=W_c=34.5$ the ratio 
$\langle \delta^2 N \rangle/\langle N \rangle$
is almost insensitive to~$L$. This justifies again the existence of
the intermediate {\em scale-invariant}
statistics at the long-range energy correlations. The ratio  
$\langle \delta^2 N \rangle/\langle N \rangle$ at the MIT
decreases very slowly with increasing
the mean level number~$\langle N \rangle$.

Recently the analytical theory has been proposed~\cite{Kravtsov96} 
suggesting that the compressibility at the mobility edge is totally
characterised by properties of the critical eigenstates and is directly
determined in terms of the multifractal exponent $\mu=d-D_2$
\begin{equation}
\kappa_{c}=\lim_{\langle N \rangle \to \infty}
\frac{d\langle \delta^2 N \rangle}{d \langle N \rangle}
=\frac{\mu}{2d}<\frac{1}{2}.
\label{compr}
\end{equation} 
The results of $\mu$ and $\kappa_c$ obtained numerically so far 
for lower dimensions $d \le 3$ satisfy reasonably this relation. 
Our 4D-data for sufficiently large values 
$\langle N \rangle > 20$ yet
obey at least the condition for the above upper limit, going down to 
$\kappa_{c}\approx 0.45-0.5$. This value appears to be larger than that
of 3D, indicating that the critical spectral rigidity diminishes.
However the accuracy of $\langle \delta^2 N \rangle$ is still not
high enough to reach a precise saturation value 
at $\langle N \rangle \gg 1$
and, as a result, to provide a reliable estimate of $\mu$ 
from~(\ref{compr}). In addition, an independent computational
analysis of the multifractality of eigenfunctions in 4D would be needed.

\section{Conclusions}
We have numerically calculated the critical distribution
of neighbouring spacings $P(s)$ at the metal-insulator transition
for the 4D Anderson model. The disorder-induced crossover
between the Wigner and the Poisson statistics for finite systems 
is shown to obey a one-parameter scaling law as for lower dimensions. 
The finite size scaling analysis of the level statistics allows one 
to locate 
the critical disorder $W_{c}$
and to detect the correlation length exponent $\nu$.
Comparing the obtained results with lower 
dimensions 
we are led to the conclusion
 that the spectral correlations at criticality 
{\em depends} on the spatial dimensionality,
becoming weaker with increasing~$d$. The further systematic study of the
dimensionality dependence would be desirable in order to 
answer the question
how the {\em critical} level statistics favours the Poissonian limit,
when approaching the upper bound of the Anderson 
transition~\cite{Mirlin94}.
Another interesting problem for simulations in 4D could be spectral 
fluctuations in a weakly disordered metal to check the non-perturbative 
theory~\cite{AltshulerAA95}.
    \vspace{0.6cm}\newline{\small 
\section{Acknowledgements}
We thank  L. Schweitzer, A.D. Mirlin and M. Schreiber 
for useful discussions. 
I.Kh.Zh. thanks DFG for the financial 
support during his stay at the University of Hamburg. 
The support from the TMR-Network (contract FMRX CT96-0042) 
and the  
Sonderforschungbereich 508 ``Quantenmaterialien'' of the University
of Hamburg is gratefully acknowledged. 
    }
    \end{document}